\documentclass[conference,a4paper]{IEEEtran}

\IEEEoverridecommandlockouts

\usepackage{graphicx,array}
\usepackage{amssymb,amsmath}
\usepackage{cite}

\def\r{\mathbb{R}}
\def\vp{\varphi}

\title{Cycle slipping in nonlinear circuits under periodic nonlinearities and time delays.}

\author{ Vera Smirnova$^{1}$,  Anton Proskurnikov$^{2}$, and Natalia V. Utina$^{3}$
\thanks{*The paper is supported by RSF, grant 14-29-00142}
\thanks{$^{1}$Vera Smirnova is with the Department of Mathematics,
      St.Petersburg State University of Architecture and Civil Engineering,
      the Department of Mathematics and Mechanics,
      St.Petersburg State University,
      St.Petersburg, Russia,
      {\tt\small root@al2189.spb.edu}}%
\thanks{$^{2}$Anton Proskurnikov is with the Research Institute of Technology and Management, University of Groningen. He is also with
      St.Petersburg State University, ITMO University, and Institute for Problems of Mechanical Engineering RAS,
      St.Petersburg, Russia
      {\tt\small avp1982@gmail.com}}%
\thanks{$^{3}$ Natalia Utina is with the Department of Mathematics,
      St.Petersburg State University of Architecture and Civil Engineering,
      St.Petersburg, Russia,
      {\tt\small unv74@mail.ru}}%

}

\begin{document}

\maketitle

\begin{abstract}
Phase-locked loops (PLL), Costas loops and other synchronizing circuits are featured by the presence of a nonlinear \emph{phase detector}, described by a periodic nonlinearity. In general, nonlinearities can cause complex behavior of the system such multi-stability and chaos. However,
even phase locking may be guaranteed under any initial conditions, the transient behavior of the circuit can be unsatisfactory due to the \emph{cycle slipping}. 
Growth of the phase error caused by cycle slipping is undesirable, leading e.g. to demodulation and decoding errors. This makes the problem of estimating the 
phase error oscillations and number of slipped cycles in nonlinear PLL-based circuits extremely important for modern telecommunications. 
Most mathematical results in this direction, available in the literature, examine the probability density of the phase error and expected number of slipped cycles under
stochastic noise in the signal. At the same time, cycle slipping occurs also in deterministic systems with periodic nonlinearities, depending on the initial conditions, properties of the linear part and the periodic nonlinearity and other factors such as delays in the loop. In the present paper we give analytic estimates for the number of slipped cycles in PLL-based systems,
governed by integro-differential equations, allowing to capture effects of high-order dynamics, discrete and distributed delays. We also consider the effects of singular small-parameter perturbations
on the cycle slipping behavior.
\end{abstract}


\section{Introduction}

A lot of systems arising in electrical engineering, industrial electronics and telecommunications are based on the seminal idea of the phase-locked loop (PLL) and contain a digital or analog
circuits, which synchronizes some internal oscillator with an exogenous periodic signal in phase (which task is sometimes referred to as \emph{phase tracking} or \emph{phase recovery}) \cite{Gardner,Lindsey,Margaris}. Mathematical model of a PLL-based circuit typically may be considered as a feedback interconnection of a linear time-invariant system and a periodic nonlinearity, characterizing the \emph{phase detector}. Such systems have been widely studied in mechanics and control theory under different names (``phase synchronization systems'', ``pendulum-like systems'' etc.), see e.g. \cite{Leonov:06,LeonovReitmannSmirnova} and references therein.

Under natural assumptions it is often possible to prove the steady-state phase locking (that is, the phase error converges to one of the equilibria), which sometimes is called ``gradient-like behavior'', see \cite{Leonov:06,LeonovReitmannSmirnova} and references therein. Before achieving the synchronous regime, the phase error normally oscillates around some equilibrium point.
However, due to large fluctuations it may leave the corresponding region of attraction and be attracted by another equilibrium point. During this transition, known as the \emph{cycle slipping},  
the phase shift significantly increases which is especially undesirable in data transmission systems, resulting in demodulation errors.

Appealing to a simple mechanical analogue of the PLL-based circuit, that is, the model of mathematical pendulum, the process without cycle slipping corresponds to the oscillation around the stable lower equilibrium, whereas the cycle slipping is portrayed by the passing via the upper unstable point. Starting from this basic model, J. Stoker \cite{Stoker} suggested a definition for the number of slipped cycles, being a crucial characteristics of the transient process in a PLL circuit. Given a $\Delta$ - periodic solution with the phase error $\sigma(t)$, the \emph{number of slipped cycles} is defined by $k:=\max_{t\ge 0}\lfloor |\sigma(t)-\sigma(0)|/\Delta\rfloor$. In other words, $|\sigma(t)- \sigma(0)| < (k+1)\Delta$ for any $t\ge 0$, however $|\sigma(\widehat{t})- \sigma(0)| = k \Delta$ for some $\widehat{t}\ge 0$.

Under persistent stochastic noise in the input signal, PLL exhibit random cycle slippings, whose statistical properties (such as the expected rate, average time of the first slip etc.)
have been subject of extensive studies for more than 50 years since the pioneering paper by Viterbi \cite{Viterbi:63}, see \cite{Ascheid:82},\cite{Moeneclaey:85}. 
In the present paper, we are interested in the deterministic model of cycle slipping which is caused not by external excitations but only by the nonlinear structure of the circuit itself.
We consider a very general mathematical model, described by integro-differential Volterra equations and encompassing a large class of PLL-based systems, including those with delays \cite{Bergmans:95}. The offered analytic criterion, providing the upper bound for the number of cycles slipped, is based on the Popov method for the absolute stability elaborated in control theory \cite{Rasvan:06}, involves the initial conditions, transfer function of the linear part of the system and the periodic characteristics of the phase detector. Assumptions about the solution boil down to a frequency-domain inequality and some algebraic condition. The first ``frequency-algebraic'' criterion of such a type was obtained in \cite{Ershova:83} (and later was reformulated in \cite{YangHuang:07}) and then extended to discrete-time PLL \cite{Smirnova:05} and infinite-dimensional systems \cite{LeonovReitmannSmirnova}. In the later papers \cite{Perkin:12, Perkin:14-2}
the restrictions, adopted in the mentioned previous papers, were significantly relaxed under assumption of smooth periodic nonlinearity with known a priori derivative bounds.

In this paper, we simplify and extend the results from \cite{Perkin:14-2}, and illustrate their potential by estimating the number of slipped cycles for a delayed PLL.  We also discuss robustness of our estimates against singular perturbation, introducing a higher derivative term with a small parameter (the term ``singular'' highlights that the order of unperturbed equation is lower than the order of the perturbed one). Singularly perturbed systems describe a wide range of physical, mechanical and electrical systems \cite{Nayfeh}, and their asymptotic properties and transient dynamics in general significantly differ from those of unperturbed  ones, requiring a special theory \cite{Imanaliev}.

\section{Problem setup}

We consider a nonlinear system with a periodic nonlinearity, governed by an integro-differential Volterra equation
\begin{equation}
 \begin{array}{l}
   \dot \sigma (t) = \alpha(t)+ \rho \varphi (\sigma (t-h)) 
 -\int \limits_0^{t} \gamma(t-\tau)\varphi((\sigma (\tau)) \,d\tau.
                                                                                                                 \label{4}
\end{array}
\end{equation}

Here ${h\ge 0}$, ${\rho \in{\bf R}}$, ${\gamma, \alpha : [0,+\infty)\to {\bf R}}$. The map ${\varphi:{\bf R}\to {\bf R}}$ is assumed ${C^1}$-smooth and ${\Delta}$-periodic with two simple isolated roots on ${[0,\Delta)}$. The kernel function ${\gamma(\cdot)}$ is piece-wise continuous, the function ${\alpha(\cdot)}$ is continuous.
The solution of~(\ref{4}) is defined uniquely by the initial conditions
\begin{equation}
\sigma(t)|_{t\in [-h,0]} = \sigma^0(t),\quad \sigma(0+0)=\sigma^0(0),                                                                                                            \label{eqn_111}
\end{equation}
where ${\sigma^0(\cdot)}$ is a continuous function. The system \eqref{4} may be considered as a feedback interconnection of a linear system, whose transfer function
is defined as
\begin{equation}
   K(p) = -\rho e^{-hp}  + \int \limits_0^{t} \gamma(t) e^{-pt} \,dt \ \ \ (p \in {\bf C}),
                                                                                                                                      \label{5}
\end{equation}
and a nonlinear block $\sigma\mapsto\vp(\sigma)$. Throughout the paper, we assume the linear part of \eqref{4} is exponentially stable, that is, 
$$
|\alpha(t)| + |\gamma(t)| \le Me^{-rt} \ \ \ (M,r > 0).
$$
For definiteness, we assume that 
\begin{equation}
  \int \limits_0^\Delta \varphi(\sigma) \,d\sigma  \le 0.             \label{3}
\end{equation}
We assume the lower and upper slopes $\alpha_1,\alpha_2$ for the periodic nonlinearity $\vp(\cdot)$ are known:
$$
\alpha_1: = \inf\limits_{\sigma \in [0,\Delta)} \frac{d\varphi}{d\sigma}  < 0 < \alpha_2 := \sup\limits_{\sigma \in [0,\Delta)} \frac{d\varphi}{d\sigma}.
$$

Our goal is to disclose the maximal phase error deviation  ${ \sup\limits_{t\ge 0} |\sigma(t) - \sigma(0)|}$ along the solution, which also gives the upper bound for the number
of slipped cycles.

\section{Phase error estimates}

The main idea of the phase error estimation, proposed in \cite{LeonovReitmannSmirnova,Perkin:14-2} is a modified Popov's method of ``a priori integral indices'' which was elaborated to prove
absolute stability of nonlinear systems \cite{Rasvan:06}. The cornerstone of this method is an integral quadratic constraint, which is guaranteed by a frequency-domain inequality.

{\bf Lemma 1. \cite{Perkin:14-2}} {\it
Suppose there exist such positive
$\vartheta$, $\varepsilon$, $\delta$, $\tau$
that
for all $\omega\geq 0$ the frequency-domain inequality holds:
\begin{equation}\label{eqn_100}
\begin{array}{l}
Re \{ \vartheta K(i\omega)- \tau(K(i\omega)+\alpha_1^{-1}i\omega)^*(K(i\omega)+\alpha_2^{-1}i\omega)\}  \\
- \varepsilon |K(i\omega)|^2-\delta \geq 0 \quad (i^2=-1).
\end{array}
\end{equation}
\noindent
Then the following integral quadratic functionals
$$
\begin{array}{c}
I_T[\sigma(\cdot)]=\displaystyle\int\limits_0^T \left\{\vartheta\dot{\sigma}(t)
\varphi(\sigma(t))+\varepsilon\dot{\sigma}^2(t) +\delta\varphi^2(\sigma(t))+\right.\\
\left.+\tau(\alpha_1^{-1}\dot{\varphi}(\sigma(t))-\dot{\sigma}(t))
(\alpha_2^{-1}\dot{\varphi}(\sigma(t))-\dot{\sigma}(t))\right\}dt
\end{array}
$$
are uniformly bounded along the solution
of (\ref{4}):
\begin{equation} \label{eqn_101}
I_T\leq Q,
\end{equation}
where $Q$ does not depend on $T$.
}

A closer analysis of the proof in \cite{Perkin:14-2} shows that the value of $Q$ in fact depends on the parameters $\vartheta$, $\varepsilon$, $\delta$, $\tau$, the dynamics of linear part of \eqref{4} and also the value of $\max|\vp(\sigma(t))|$, which in practice always can be estimated from the equations \eqref{4}. However, this estimates appears to be rather involved and conservative.
Assuming that some estimate $Q$ is known, the following two estimates for the number of slipped cycles main be obtained \cite{Perkin:14-2}.

To start with, we introduce the following auxiliary functions
\begin{gather*}
\Phi(\sigma) := \sqrt{(1-\alpha_1^{-1}\varphi'(\sigma))(1-\alpha_2^{-1}\varphi'(\sigma))}\\
P(\varepsilon, \tau,\sigma) := \sqrt{\varepsilon +\tau\Phi^2(\sigma)}\\
r_j(k, \vartheta, x) := \frac{\int\limits_0^\Delta\varphi(\sigma)d\sigma
+ (-1)^j\frac{x}{\vartheta k}}{\int\limits_0^\Delta
|\varphi(\sigma)|d\sigma},\\
r_{0j}(k, \vartheta, x) :=
\frac{\int\limits_0^\Delta\varphi(\sigma)d\sigma +
(-1)^j\frac{x}{\vartheta k}}{\int\limits_0^\Delta \Phi(\sigma)
|\varphi(\sigma)|d\sigma},\\
r_{1j}(k, \vartheta, \varepsilon, \tau, x) :=
\frac{\int\limits_0^\Delta\varphi(\sigma)d\sigma +
(-1)^j\frac{x}{\vartheta k}}{\int\limits_0^\Delta
|\varphi(\sigma)| P(\varepsilon, \tau,\sigma) d\sigma} \quad (j=1,2)\\
Y_j(\sigma): =
\varphi(\sigma)- r_{1j}|\varphi(\sigma)|P(\varepsilon, \tau,\sigma) \quad  (j=1,2)
\end{gather*}
and also a matrix-valued function
$ T_j(k, \vartheta, x, a):=$
\[
:=\left\|
\begin{array}{ccccc}
\varepsilon & \displaystyle\frac{a\vartheta r_j(k,\vartheta,x)}{2} & 0 \\
\displaystyle\frac{a\vartheta r_j(k,\vartheta,x)}{2} &  \delta &
\displaystyle\frac{a_0\vartheta r_{0j}(k,\vartheta,x)}{2} \\
0 &  \displaystyle\frac{a_0\vartheta r_{0j}(k,\vartheta,x)}{2} & \tau
\end{array}
\right\|,
\]
where $a\in[0,1]$ and $a_0:=1-a$.

{\bf Theorem 1.} {\it
Suppose there exist such positive
$\vartheta$, $\varepsilon$, $\delta$, $\tau$ and
natural $k$   that the following conditions are fulfilled:

1) for all $\omega\geq 0$ the inequality (\ref{eqn_100}) is valid;

2)the condition holds
\begin{equation} \label{eqn_5}
4\delta > \vartheta^2(r_{1j}(k,\vartheta,\varepsilon, \tau, Q))^2  \quad  (j=1,2),
\end{equation}
where $Q$ is the bound from (\ref{eqn_101}). Then any solution of (\ref{4}) slips less than $k$ cycles, that is, the inequalities hold
\begin{equation} \label{eqn_33}
|\sigma(0)-\sigma(t)|<k\Delta\quad\forall t\ge 0.
\end{equation}
}

The conditions of Theorem~1 resemble those from \cite{Ershova:83}, however, they are applicable for general infinite-dimensional system \eqref{4} and visibly improve
the result from \cite{Ershova:83} even for the case of ordinary differential equations \cite{Perkin:12}.

{\bf Theorem 2.} {\it
Suppose there exist positive
$\vartheta$, $\varepsilon$, $\delta$, $\tau$, $a \in [0,1] $ and
natural $k$ satisfying the conditions as follows:

1) for all $\omega\geq 0$ the inequality (\ref{eqn_100}) is valid;

2) the matrices $T_j(k,\vartheta,Q,a)$ $(j=1,2)$  where the value of $Q$
is defined by (\ref{eqn_101}),
are positive definite.

\noindent
Then for the solution of (\ref{4}) the inequality (\ref{eqn_33}) holds.
}

Conditions of the latter theorem may be seriously simplified in a special case where $\alpha_1=-\alpha_2$ and $\vp(\sigma(0))=0$. Retracing the estimates from \cite{Perkin:14-2}, one can show that  $$Q\leq q:=\frac{1}{r}\left(\vartheta Mm+2(\varepsilon+\tau)Mm(\frac{M}{r}+\rho)+(\varepsilon+\tau)\frac{M^2}{2}\right),$$
where
$
m:=\sup{\varphi(\sigma)}.
$
This gives rise to the following simplified version of Theorem~2.

{\bf Theorem 3} {\it Let $\alpha_1=-\alpha_2$. 
Suppose there exist such positive
$\vartheta$, $\varepsilon$, $\delta$, $\tau$, $a \in [0,1] $ and
natural $k$   that the following conditions are fulfilled:

1) for all $\omega\geq 0$ the frequency-domain inequality(\ref{eqn_100}) holds;

2) the matrices $T_j(k,\vartheta,q)$ $(j=1,2)$
are positive definite.

\noindent
If $\sigma(0)=\sigma_0$ where $\varphi(\sigma_0)=0$, then for any solution of (\ref{4}) the estimate (\ref{eqn_33}) holds.
In general, the estimate (\ref{eqn_33}) is valid, replacing $k$ with $k+1$. 
}

{\bf Theorem 3} {\it
Let $\sigma(0)=\sigma_0$ where $\varphi(\sigma_0)=0$.
Suppose there exist such positive
$\vartheta$, $\varepsilon$, $\delta$, $\tau$, $a \in [0,1] $ and
natural $k$   that the following conditions are fulfilled:

1) for all $\omega\geq 0$ the frequency-domain inequality(\ref{eqn_100}) holds;

2) the matrices $T_j(k,\vartheta,q)$ $(j=1,2)$
are positive definite.

\noindent
Then for any solution of (\ref{4}) the estimate (\ref{eqn_33}) holds.
}

The first claim of Theorem~3 follows from Theorem~2, since, as was mentioned, $q$ may be used instead of $Q$ under restriction $\varphi(\sigma_0)=0$. To prove the second claim,
notice that $\vp(\sigma)$ is $\Delta$-periodic and has zeros on the period. Therefore, if $\vp(\sigma(t))\ne 0$ for any $t\ge 0$, then $|\vp(0)-\vp(t)|<\Delta$ for any $t\ge 0$, so 
the solution slips no cycles. Otherwise, let $t_0\ge 0$ be the minimal number such that $\vp(\sigma(t_0))\ne 0$. It is obvious that $|\sigma(t_0)-\sigma(0)|<\Delta$. Applying Theorem~2 for the solution $\sigma(t-t_0)$, one shows that $|\sigma(t_0)-\sigma(t)|<k\Delta$ for any $t\ge t_0$ and hence $|\sigma(0)-\sigma(t)|<(k+1)\Delta$.

\section{Example}

Let us consider a phase-locked loop (PLL) with a proportional integral low-pass filter, a sine-shaped characteristic of phase frequency detector and a time-delay in the loop.
Its mathematical description is borrowed from \cite{Belustina}:
\begin{equation} \label{eqno_307}
\ddot\sigma(t)+\frac{1}{T}\dot\sigma(t)+\varphi(\sigma(t-h))+sT\dot\varphi(\sigma(t-h))=0,
\end{equation}
$$
\varphi(\sigma)=\sin\sigma -\beta, \, s \in (0,1), \,  \beta \in (0,1], \, h>0, \, T>0.
$$
The differential equation (\ref{eqno_307}) can be reduced to integro-differential equation  (\ref{4})
with
\begin{equation*}
\gamma(t) = \left\{
                   \begin{array}{ccl}
                     0, & t<h, \\
                    (1-s)e^{-\frac{t-h}{T}}, & t\geq h \\
                   \end{array}
\right\},
\end{equation*}
\begin{equation*}
\alpha(t) =e^{-\frac{t}{T}}(b-(1-s)J),
\end{equation*}
where $b=\dot\sigma(0)+sT\varphi(\sigma(-h))$ and
\begin{equation*}
J =
\left
\{
\begin{array}{ccl}
      \displaystyle
\int\limits_{-h}^{t-h}e^{\frac{\lambda+h}{T}}\varphi(\sigma(\lambda))d\lambda, & t\leq h, \\
       \displaystyle
\int\limits_{-h}^{0}e^{\frac{\lambda+h}{T}}\varphi(\sigma(\lambda))d\lambda, & t> h
\end{array}
\right\}.
\end{equation*}

The transfer function of the lowpass filter here has the form:
\begin{equation*}
K(p)=T\frac{Tsp+1}{Tp+1}e^{-ph}
\end{equation*}

We suppose that $\varphi(\sigma(0))=0$ and apply Theorem 4.

Let $\alpha_2=-\alpha_1=1,\,\vartheta=1, \, a=1$.
The assumption 1) of  Theorem 4 shapes into
\begin{equation} \label{eqno_308}
\begin{array}{ll}
\Omega(\omega)\equiv \tau T^2\omega^4 + \omega^2(T^3s\cos{\omega h}-T^4s^2(\varepsilon+\tau) +\\
+\tau-\delta T^2) -T^2(1-s)\omega\sin{\omega h} +T\cos{\omega h} - \\
-(\varepsilon+\tau)T^2-\delta \geq 0 \quad  \forall \omega;
\end{array}
\end{equation}
whereas condition 2) may be rewritten as
\begin{equation} \label{eqno_309}
2\sqrt{\varepsilon\delta} > \frac{2\pi\beta+q_2k^{-1}}{4(\beta\arcsin\beta+\sqrt{1-\beta^2})}.
\end{equation}

Notice that for all $\omega\in\r$ one has
\begin{equation*}
\begin{array}{ll}
\Omega(\omega)\geq \Omega_0(\omega)\equiv (\tau T^2-\frac{1}{2}T^3sh^2)\omega^4 + (T^3s-\\
-T^4s^2(\varepsilon+\tau) +\tau-\delta T^2-\frac{1}{2}Th^2 -(1-s)T^2h)\omega^2+\\
+(T-(\varepsilon+\tau)T^2-\delta), \quad \forall \omega
\end{array}
\end{equation*}
and $\Omega(\omega)\approx\Omega_0(\omega)$ when $\omega h<<1$.

We consider the case $T\leq 0.9,\, h_0= \frac{h}{T}\leq 1$, since for small $T$ and small $h$ the PLL is
gradient-like for all $\beta \in (0,1]$ \cite{Belustina}.
Let us choose $\varepsilon=\frac{\beta_0}{T},\, \delta=\alpha_0T,\, \tau=\gamma_0 T^3$.
As $\Omega(0)=\Omega_0(0)$ it is necessary that $\alpha_0+\beta_0+\gamma_0T^4\leq 1$.
Then the optimal values for $\alpha_0$ and $\beta_0$ are $\alpha_0=\beta_0=\frac{1}{2}(1-\gamma_0T^4)$,
whence $2\sqrt{\varepsilon\delta}=1-\gamma_0T^4$.
For $\gamma_0=\max{\{\frac{1}{2}sh_0^2,\, \frac{1}{2}(h_0+1-s)^2\}}$ the polynomial   $\Omega_0(\omega)$
is nonnegative, $\forall\omega$.
It follows from (\ref{eqno_309}) that the number $k_0$ of cycles slipped satisfies the inequality
\begin{equation*}
k_0\leq r_0:=\lfloor q_2(8\sqrt{\varepsilon\delta}(\beta\arcsin\beta+\sqrt{1-\beta^2})-2\pi\beta)^{-1}\rfloor.
\end{equation*}

Let us consider the PLL with $b=K(0)\beta$ \cite{Ershova:83}.
Then by estimating the functional $I_T$ we conclude that the value of $q$ can be defined by the
formula
\begin{equation} \label{eqno_310}
q=T^2(A+Bh_0+Ch_0^2),
\end{equation}
where
\begin{equation} \label{eqno_311}
\begin{array}{ll}
A=(\frac{7}{2}\beta^2 +3), \\
B=3(1-s)(1+\beta)(3\beta+1),\\
C=\frac{3}{2}(1-s)^2(1+\beta)^2.
\end{array}
\end{equation}
It follows from (\ref{eqno_310}), (\ref{eqno_311}) that the number of slipped cycles increases together
with $T$, with $\beta$ or with $h_0$.
Let for example $h_0=1,\, s=0.4,\,T=0.1$.
Then $r_0=1$ for $\beta=0.9$,\, $r_0=2$ for $\beta=0.92$, and $r_0=5$ for $\beta=0.95$.

\section{Extension: robustness to small singular perturbations}

The estimates for the number of slipped cycles presented in the paper are easily shown to be robust against small variations in the parameters of the transfer function and the nonlinearity.
It appears, however, that they remain robust even to a singular perturbation, introducing a high-order term into equation \eqref{4}:

\begin{equation}
 \begin{array}{l}
   \mu \ddot \sigma_\mu (t) + \dot \sigma_\mu (t) = \alpha(t)+ \rho \varphi (\sigma_\mu (t-h))- \\
 -\int \limits_0^{t} \gamma(t-\tau)\varphi((\sigma_\mu (\tau)) \,d\tau \ \ \ (t \ge 0).
                                                                                                                                   \label{1}
\end{array}
\end{equation}
Here ${\mu>0}$ is a small parameter. Equation (\ref{1}) can be reduced to the form \eqref{4}
\begin{equation} \label{eqn_51}
\begin{array}{l}
\dot{\sigma_\mu}(t)=\alpha_\mu(t)
-\displaystyle\int\limits_0^t\gamma_\mu(t-\tau)\varphi(\sigma_\mu(\tau))d\tau
\quad (t>0),
\end{array}
\end{equation}
where
$\alpha_\mu(t)=\displaystyle\dot{\sigma}(0)e^{\frac{-t}{\mu}}+\frac{1}{\mu}
\int\limits_{0}^{t}e^{\frac{\lambda-t}{\mu}}\alpha(\lambda)d\lambda +
\frac{\rho}{\mu}J_0(t)$,
\begin{equation*}
J_0(t) =
\left
\{
\begin{array}{ccl}
      \displaystyle
\int\limits_{-h}^{t-h}e^{\frac{\lambda+h-t}{\mu}}\varphi(\sigma(\lambda))d\lambda, & t\leq h, \\
       \displaystyle
\int\limits_{-h}^{0}e^{\frac{\lambda+h-t}{\mu}}\varphi(\sigma(\lambda))d\lambda, & t> h,
\end{array}
\right\}.
\end{equation*}
\begin{equation} \label{eqn_53}
\gamma_\mu(t)=\displaystyle\frac{1}{\mu}
\int\limits_{0}^{t}e^{\frac{\lambda-t}{\mu}}\gamma(\lambda)d\lambda -
\frac{\rho}{\mu}
\left
\{
\begin{array}{ccl}
e^{\frac{h-t}{\mu}}, &  t\geq h,\\
0 , & t< h
\end{array}
\right \}.
\end{equation}

The transfer function for equation (\ref{eqn_51}) is given by
\begin{equation} \label{eqn_54}
K_\mu(p)=\frac{K(p)}{1+\mu p}.
\end{equation}
Let $
q_0=q+(\vartheta Mm+2(\varepsilon+\tau)Mm(\frac{M}{r}+\rho)\rho mh +(\varepsilon+\tau) \rho^2 m^2 h.
$
Applying Theorem~3 to \eqref{eqn_51}, one gets the following.

{\bf Theorem 4} {\it
Let $\alpha_2=-\alpha_1$ and
$\sigma(0)=\sigma_0$ where $\varphi(\sigma_0)=0$.
Suppose there exist such positive
$\vartheta$, $\varepsilon$, $\delta$, $\tau$, $a \in [0,1] $ and
natural $k$   that the following conditions are fulfilled:

1) for all $\omega\geq 0$ the frequency-domain inequality(\ref{eqn_100}) holds;

2) the matrices $T_j(k,\vartheta,q_0)$ $(j=1,2)$
are positive definite.

\noindent
Then there exists such value $\mu_0$ that  for all $\mu \in (0,\mu_0)$ and any solution of (\ref{1}) the estimate (\ref{eqn_33}) holds.
}

\section{Conclusion}

We consider the problem of cycle slipping for PLL based circuits, governed by integro-differential Volterra equations, which model captures, in particular, potential delays in the system.
Under restriction of smooth nonlinearity, we get analytic estimates for the number of slipped cycles. We also consider small singular perturbations of our equations, showing
the estimates are uniform with respect to a small parameter.

\bibliographystyle{plain}
\bibliography{literature}

\begin{thebibliography}{10}

\bibitem{Ascheid:82}
G.~Ascheid and H.~Meyr.
\newblock Cycle slips in phase-locked loops: A tutorial survey.
\newblock {\em IEEE Trans. on Commun.}, 30(10):2228--2241, 1982.

\bibitem{Belustina}
L.N. Belustina.
\newblock Qualitative-numerical analysis for synchronization system with
  time-delay (in russian).
\newblock In {\em Proceedings of International Seminar "Nonlinear circuits and
  systems"}, volume~2, pages 50--59, Wien, 1992. Springer-Verlag.

\bibitem{Bergmans:95}
J.W.M. Bergmans.
\newblock Effect of loop delay on stability of discrete-time {PLL}.
\newblock {\em IEEE Trans. Circuits Syst. I}, 42(4):229–--231, 1995.

\bibitem{Ershova:83}
O.~B. Ershova and G.~A. Leonov.
\newblock Frequency estimates of the number of cycle slippings in the phase
  control systems.
\newblock {\em Autom. Remote Control}, 44(5):600--607, 1983.

\bibitem{Gardner}
F.M. Gardner.
\newblock {\em Phaselock Techniques}.
\newblock Wiley, New York, 1966.

\bibitem{Imanaliev}
M.~Imanaliev.
\newblock {\em Oscillation and stability of solutins of singularly-perturbed
  integro-differential systems}.
\newblock ILIM, Frunze, 1974.

\bibitem{LeonovReitmannSmirnova}
G.~A. Leonov, V.~Reitmann, and V.~B. Smirnova.
\newblock {\em Non-local methods for pendulum-like feedback systems}.
\newblock Teubner, Stuttgart-Leipzig, 1992.

\bibitem{Leonov:06}
G.A. Leonov.
\newblock Phase synchronization. theory and applications.
\newblock {\em Autom. Remote Control}, 67(10):1573--1609, 2006.

\bibitem{Lindsey}
W.C. Lindsey.
\newblock {\em Synchronization Systems in Communication and Control}.
\newblock Prentice-Hall, Englewood Cliffs, NJ, 1972.

\bibitem{Margaris}
W.~Margaris.
\newblock {\em Theory of the Non-linear Analog Phase Locked Loop}.
\newblock Springer, New York, 2004.

\bibitem{Moeneclaey:85}
M.~Moeneclaey.
\newblock The influence of phase-dependent loop noise on the cycle slipping of
  symbol synchronizers.
\newblock {\em IEEE Trans. on Commun.}, 33(12):1234--1239, 1985.

\bibitem{Nayfeh}
A.H. Nayfeh and D.T. Mook.
\newblock {\em Nonlinear oscillations}.
\newblock WILEY-VCH Verlag GmbH \& Co, 2004.

\bibitem{Perkin:14-2}
A.~Perkin, A.~Proskurnikov, and V.~Smirnova.
\newblock Asymptotic estimates for gradient-like distributed parameter systems
  with periodic nonlinearities.
\newblock In {\em Proceedings of 2014 IEEE Multi-conference on Systems and
  Control (MSC 2014)}, pages 1638--1643, Antibes, France, 2014.

\bibitem{Perkin:12}
A.A. Perkin, V.B. Smirnova, and A.I. Shepeljavyi.
\newblock Frequency-algebraic conditions for stability of phase systems with
  application to phase-locked loops and synchronization systems.
\newblock {\em Cybernetics and Physics}, 1(3):188--197, 2012.

\bibitem{Rasvan:06}
V.~Rasvan.
\newblock Four lectures on stability. {L}ecture 3. {T}he frequency domain
  criterion for absolute stability.
\newblock {\em Control Engineering and Applied Informatics}, 8(2):13--20, 2006.

\bibitem{Smirnova:05}
A.~I. Shepeljavyi, V.~B. Smirnova, and N.~V. Utina.
\newblock Frequency-domain estimates for transient attributes of discrete phase
  systems.
\newblock In {\em Proceedings of International Conference on Physics and
  Control (PhysCon 2005)}, pages 469--473, St. Petersburg, Russia, 2005.

\bibitem{Stoker}
J.J. Stoker.
\newblock {\em Nonlinear vibrations in mechanical and electrical systems}.
\newblock Interscience, New York, 1950.

\bibitem{Viterbi:63}
A.J. Viterbi.
\newblock Phase-locked loop dynamics in the presence of noise by
  {F}okker-{P}lanck techniques.
\newblock {\em Proceedings of IEEE}, 51:1737--1753, 1963.

\bibitem{YangHuang:07}
Y.~Yang and L.~Huang.
\newblock Cycle slipping in phase synchronization systems.
\newblock {\em Phys. Lett. A}, 362:183--188, 2007.

\end{thebibliography}

\end{document}